\documentstyle[12pt]{article}

\newcommand{\beq}{\begin{equation}}
\newcommand{\eeq}{\end{equation}}
\newcommand{\bea}{\begin{eqnarray}}
\newcommand{\eea}{\end{eqnarray}}
\newcommand{\half}{\frac{1}{2}}
\newcommand{\ihalf}{\frac{i}{2}}

\newcommand{\opsi}{\overline{\Psi}}

\newcommand{\opar}{\overline{\epsilon}}

\newcommand{\dslash}{\not\!\! D}

\newcommand{\dboth}{\stackrel{\leftrightarrow}{D}}
\newcommand{\dmu}{\partial_{\mu}}

\begin{document}
\title
{Bogomol'nyi Bounds and Killing Spinors  in $d=3$ Supergravity}
\author{
Jos\'e D. Edelstein$^a$\thanks{CONICET}~,
Carlos N\'u\~nez$^a$ and
Fidel A. Schaposnik$^b$\thanks{Investigador CICBA, Argentina}
\thanks{On leave from Universidad Nacional de La Plata, Argentina}\\
{}~
\\
$^a$Departamento de F\'\i sica, Universidad Nacional de La Plata\\
C.C. 67, 1900 La Plata, Argentina\\
{}~\\
$^b$Laboratoire de Physique Th\'eorique ENSLAPP
\thanks{URA 1436 du CNRS associ\'e
\`a l'Ecole Normale Sup\'erieure de Lyon et \`a
l'Universit\'e de Savoie,}\\
B.P. 110, F-74941 Annecy-le-Vieux, France}
\date{\today}
\maketitle
\thispagestyle{empty}

\def\thepage{\protect\raisebox{0ex}{\ } La Plata-Th 95/32}
\thispagestyle{headings}
\markright{\thepage}

\begin{abstract}
We discuss the connection between the construction
of Bo\-go\-mol'\-nyi bounds and equations
in three dimensional gravitational theories and
the existence
of an underlying $N=2$ local supersymmetric structure.
We show that, appart from matter self duality equations,
a first order
equation for the gravitational field
(whose consistency condition gives the
Einstein equation) can be written
as a consequence of the
local supersymmetry. Its solvability makes possible
the evasion of the
no-go scenario for the construction of Killing spinors in
asymptotically
conical spacetimes. In particular we show that
the existence of non-trivial supercovariantly
constant spinors
is guaranteed whenever field configurations saturate the topological
bound.
\end{abstract}

\newpage
\setcounter{page}{1}
\pagenumbering{arabic}

The connection between Bogomol'nyi (topological) bounds \cite{Bogo}
for the energy or the action of classical solutions in
bosonic field theories and the $N=2$
supersymmetry algebra of the corresponding supersymmetric
extension is by now well understood. The topological charge
of the purely bosonic theory coincides with the central charge
of the supersymmetry algebra in the $N=2$ extended model
\cite{WO,HS,ENS}. The Bogomol'nyi bound then arises from
positivity of the squared supersymmetry charge algebra.
Moreover, supersymmetry of physical states leads to
Bogomol'nyi equations saturating the bound (first order
self-duality equations for the bosonic
theory). Half of the supersymmetries are broken on field
configurations that solve the Bogomol'nyi equations. The
necessary conditions for the $N=2$ supersymmetric extension,
coincide with those ensuring the existence
of Bogomol'nyi bounds. Let us finally stress that the classical
aproximation to the mass spectrum given by these bounds
is expected to
be exact at the quantum level since supersymmetry ensures
that there are no quantum corrections.

The same scenario should hold when one includes
gravity, thus considering the case of
{\it local} supersymmetry. Indeed,
several models with self-gravitating solitons
have been studied following the approach described above
\cite{Models,BBS,ENS2} and Bogomol'nyi bounds were found
by carefully analysing the $N=2$ supergravity algebra.
As in the global case,
half of the supersymmetries are broken when the bound is saturated.
The presence of remnant unbroken supersymmetries is equivalent
to the existence of Killing spinors \cite{hull}.
It is important to stress at this point that
the local algebra posses certain problems which,
as we shall see, can be unambiguosly
solved following
the Hamiltonian formulation proposed by Deser and
Teitelboim \cite{DT,T,BCT}).

In three dimensional spacetimes
the connection between Bogomol'nyi bounds and
local supersymmetry
exhibits certain subtleties.
To see this, let us consider
a spinor $\eta$ (such that $\gamma^0\eta = \pm\eta$) which is
parallel transported around a closed curve $\Gamma$ at
infinity. Then, $\eta$ acquires a phase given by
the circulation of the spinorial connection $\omega_{\mu}^a$
\beq
\eta(x)\vert_{2\pi} = {\cal P}\exp\left(-\ihalf\oint_{\Gamma}
\omega_{\mu}^a\gamma^adx^{\mu}\right)\eta(x)\vert_0.
\label{paral}
\eeq
Here ${\cal P}$ denotes path-ordered integration. Now,
static massive solutions
to Einstein equations  correspond,
in three dimensional spacetime, to asymptotically conical
geometries so that the circulation of the spinorial
connection is proportional to the deficit angle $\delta$.
The resulting non-trivial holonomy gives raise to an ill-defined
$\eta$. But precisely these spinors
are necessary for defining supercharges
that generate unbroken supersymmetries \cite{W3}.
The impossibility of finding
such Killing spinors would then difficult
an extension of the global approach \cite{HS,ENS} to locally
supersymmetric models in $3$ dimensions.

The present work intends
to clarify these points giving general arguments
that show the conditions under which
covariantly constant Killing spinors
are well defined in a variety of three dimensional supergravity
models. As an example of our approach we
discuss at the end of our work the $d=3$  ${CP}^n$ model coupled to $N=2$.

{}~

\noindent{\bf The setting of the problem and its solution}
\vspace{0.2 cm}

Let us start by noting that
the problem  described above
concerning the proper definition of
Killing spinors is absent in
the three dimensional models that have been
already used as laboratories
to investigate the relation above mentioned \cite{BBS,ENS2,HIPT}.
In particular, this is the case of the Abelian Higgs model coupled
to $N=2$ supergravity for
which the existence of global unbroken
supercharges, for certain solitons states,
has been recently shown \cite{BBS,ENS2}.
Also in $3$ dimensions, a new class of
$(p,q)$-extended Chern-Simons Poincar\'e supergravities
has been constructed
and the existence of Killing spinors in the $(2,0)$
theory has been proven \cite{HIPT}.
In all these examples, unbroken
supersymmetries strikingly survive the coupling to supergravity,
despite the problems posed by the conical geometry.

In
the case of the $2+1$ supersymmetric Abelian Higgs model
coupled to gravity, the reasons behind this behavior
were clarified already in \cite{BBS} but one could ask
whether it obeys to
peculiarities of a particular three dimensional
model
or if it can be understood on general grounds and for other models.
To explore this last possibility we
consider a $2+1$ dimensional field theory
describing gravity coupled to { gauged} matter with
the property that a topological
charge can be defined. In fact, one can construct
in general
a topologically conserved current $J_{\mu}$ for
theories containing
a gauge potential $A^{\lambda}$:
\beq
J_{\mu} = \epsilon_{\mu\nu\lambda}\partial^{\nu}A^{\lambda}
\label{corr}
\eeq
The corresponding topological charge takes the form
\beq
T = \int_{\Sigma}J^{\mu}d\Sigma_{\mu} =
\oint_{\partial\Sigma}A_{\mu}dx^{\mu}
\label{top}
\eeq
where the integrals are performed over a space-like surface $\Sigma$
and its contour $\partial\Sigma$. Of course,
the base manifold and the gauge group should
be such as to ensure non-triviality of $T$.
The class of theories admitting
such contruction
include the model analysed in \cite{BBS,ENS2}, the ${CP}^n$ model
discussed below and others.

Since we consider the coupling of the gauge-matter
system system to $N=2$ supergravity,
the gauge potential belongs to a vector multiplet, so that
one has a
supercovariant derivative defined as
\cite{PCW}
\beq
\hat{\cal
D}_{\mu}(\omega,{\cal A}) \equiv
{\cal D}_{\mu}(\omega) + i\frac{\kappa^2}{4}{\cal A}_{\mu}
\label{newcov}
\eeq
with ${\cal A}_{\mu} = \sum{\cal A}_{\mu}^i$, being ${\cal A}_{\mu}^i$
the spin-1 component of each vector multiplet coupled to the Einstein
supermultiplet. We are denoting by ${\cal D}_{\mu}(\omega) $
the (ordinary) supergravity covariant derivative
containing the spinorial connection. By looking to
the action of the covariant derivative (\ref{newcov})
on the gravitino field
one can conclude that, in a sense, the gravitino becomes
{\it charged} \footnote{For example, in the model of
Ref.\cite{HIPT}, it corresponds to the so-called automorphism
charge appearing in $(2,0)$ Poincar\'e supergravity.}.
As we shall see, the presence of the gauge connection in the
supercovariant derivative (\ref{newcov})
is at the root of the existence of supercovariantly constant
spinors. Indeed, the Bohm-Aharonov
holonomy of ${\cal A}_{\mu}$ will cancel out
that of the spinorial connection thus eliminating
the effects of conical structure at infinity.
This is in fact what happens in the Abelian Higgs model coupled
to $N=2$ supergravity \cite{BBS,ENS2} and in Chern-Simons
formulation of $(2,0)$ Poincare supergravity \cite{HIPT}.
We argue here that this cancellation occurs whenever a
topological object satisfiying a topological bound, is present.
Note that the $A_\mu$ field
needs not to be a dynamical gauge field: one just needs
a connection entering in the supercovariant derivative
in such a way as to cancel out the
contribution of the spinorial connection .
In particular, we
shall see that this is what happens for the $CP^n$
model, where $A_\mu$ is an auxiliary field with no dynamics.

To begin with our construction, we
shall look for the supercharge
algebra of the above defined  gauge-matter theory
coupled to $N=2$ supergravity. From this, we shall see
how Bogomol'nyi bounds and Bogomol'nyi equations
can be established. Finally
we will explain how the problem posed by a correct
definition of a covariantly constant spinor
(\ref{paral}) can be circumvented.

The complex supercharge associated to supersymmetry
transformations can be written as the circulation of the
gravitino gauge field \cite{ENS2}
\beq
{\cal Q}[\epsilon] = -
\frac{2}{\kappa}\oint_{\partial\Sigma}\opar\psi_{\mu}dx^{\mu}.
\label{qsurf}
\eeq
This expression can be obtained from
dimensional reduction of
the four dimensional one obtained in Ref.\cite{T}.
Alternatively, it can be
constructed by applying the Hamiltonian approach described
in \cite{T} to the model under consideration. This last
approach needs a carefull treatment of
improper supersymmetry
transformations
\cite{BCT} which difficult the computation of the
supercharge algebra if one naively evaluates Poisson brackets.
We will discuss this issue bellow and
for the moment we simply
compute the supercharge algebra by acting
on the integrand of (\ref{qsurf}) with a supersymmetry transformation
\beq
\{\bar{{\cal Q}}[\epsilon],{\cal Q}[\epsilon]\}\vert \equiv
i\delta_{\epsilon}{\cal Q}[\epsilon]\vert =
\frac{2i}{\kappa}\oint_{\partial\Sigma}\opar\delta_{\epsilon}\psi_{\mu}\vert
dx^{\mu}.
\label{qeqe}
\eeq
For a given functional ${\cal F}$ depending both on bosonic and
fermionic fields, ${\cal F}\vert$ means ${\cal F}\vert_{\{\Psi\} = 0}$.
Being the transformation of the gravitino field given by:
\beq
\delta_{\epsilon}\psi_{\mu} =
\frac{2}{\kappa}\hat{\cal D}_{\mu}(\omega,{\cal A})\epsilon
\label{transf}
\eeq
we obtain:
\beq
\{\bar{{\cal Q}}[\epsilon],{\cal Q}[\epsilon]\}\vert =
\frac{4i}{\kappa^2}\oint_{\partial\Sigma}
\opar\hat{\cal D}_{\mu}(\omega,{\cal A})\epsilon dx^{\mu}.
\label{qeqe2}
\eeq
It is interesting to note that the above
expression is an identity
between the supercharge algebra
evaluated in the purely bosonic sector
and the circulation of the so-called
generalized Nester form
\beq
\Omega = \opar\hat{\cal D}_{\mu}(\omega,{\cal A})\epsilon dx^{\mu}.
\label{nester}
\eeq

Using the explicit expression for the covariant derivative
system given in (\ref{newcov}), eq.(\ref{qeqe2}) can be written as
\beq
\{\bar{{\cal Q}}[\epsilon],{\cal Q}[\epsilon]\}\vert =
\frac{4i}{\kappa^2}\oint_{\partial\Sigma}\opar{\cal
D}_{\mu}(\omega)\epsilon dx^{\mu}
- \oint_{\partial\Sigma}\opar\epsilon{\cal A}_{\mu}dx^{\mu}.
\label{qeqeuno}
\eeq
In order to compute these integrals, we consider the contour
of integration at large but finite radius $R$ (to avoid infrared problems),
with appropiate
asymptotic conditions on the fields. We choose by convenience
$\gamma^{0}\epsilon = \epsilon$ and an
asymptotic behaviour
\beq
\epsilon\rightarrow\Theta(R)\epsilon_{\infty}
\label{compasimp}
\eeq
where $\Theta(R)$ will be determined using the so-called
Witten condition \cite{W}.

Then, for static configurations one can  write eq.(\ref{qeqeuno})
in the following form
\beq
\{\bar{{\cal Q}}[\epsilon],{\cal Q}[\epsilon]\}\vert =
(M - \oint_{\partial\Sigma}{\cal A}_{\mu}dx^{\mu})
\opar_{\infty}\epsilon_{\infty}\Theta^{2}(R).
\label{lns}
\eeq
where we have defined $M$ (which, as we shall see
corresponds to the ADM mass)
as
\beq
M = \frac{2}{\kappa^{2}}\oint_{\partial\Sigma}\omega_{\mu}^0\gamma_0
dx^{\mu}.
\label{masa}
\eeq

We now use Stokes' theorem, to rewrite the
r.h.s. of (\ref{qeqe2}) as
\beq
\{\bar{{\cal Q}}[\epsilon],{\cal Q}[\epsilon]\}\vert =
i\int_{\Sigma}\epsilon^{\mu\nu\beta}\hat{\cal D}_{\beta}
(\opar\hat{\cal D}_{\mu}\epsilon)d\Sigma_{\nu}.
\label{cadorna}
\eeq
Then, from
\beq
[\hat{\cal D}_{\mu}(\omega,{\cal A}),\hat{\cal
D}_{\nu}(\omega,{\cal A})] =
\half {R_{\mu\nu}}^{ab}\Sigma_{ab}
+ \frac{i\kappa^2}{2}\partial_{[\mu}{\cal A}_{\nu]},
\label{nmunu}
\eeq
Einstein equations and the supersymmetry transformations
of the fermionic fields, we have
\beq
i\epsilon^{\mu\nu\beta}\hat{\cal D}_{\beta}
(\opar\hat{\cal D}_{\mu}\epsilon) = i\epsilon^{\mu\nu\beta}
\overline{\hat{\cal D}_{\beta}\epsilon}\hat{\cal D}_{\mu}\epsilon
+ \frac{\kappa^2}{2}\sum_{\{\Psi\}}
\delta_{\epsilon}\bar\Psi\gamma^{\nu}\delta_{\epsilon}\Psi
\label{divernest}
\eeq
We now specialize our spacelike integration
surface $\Sigma$ so that $d\Sigma_{\mu} =
(d\Sigma_{t},\vec{0})$. Then, we just
need to compute the time component
of eq.(\ref{divernest}) which, after some Dirac algebra, reads
\bea
i\epsilon^{t\nu\beta}\hat{\cal D}_{\beta}
(\opar\hat{\cal D}_{\mu}\epsilon) & = &
\left(\gamma^i\hat{\cal D}_i\epsilon\right)^{\dag}
\left(\gamma^j\hat{\cal D}_j\epsilon\right) -
g^{ij}\left(\hat{\cal D}_i\epsilon\right)^{\dag}
\left(\hat{\cal D}_j\epsilon\right) \nonumber \\
& + & \frac{\kappa^2}{2}\sum_{\{\Psi\}}
\delta_{\epsilon}{\bar\Psi}^{\dag}\delta_{\epsilon}\Psi
\label{compcero}
\eea
At this point, we impose the generalized Witten condition \cite{W}
on the spinorial parameter $\epsilon$
\beq
\gamma^i\hat{\cal D}_i(\omega,{\cal A})\epsilon = 0,
\label{genw}
\eeq
so that
the asymptotic behaviour of $\epsilon$ can be determined.
That is, the function $\Theta(R)$ can be computed.
On the other hand, being the r.h.s of eq.(\ref{compcero})
a sum of bilinear terms once (\ref{genw}) is applied,
positivity of the l.h.s. in eq.(\ref{cadorna})
is obtained:
\beq
\{\bar{{\cal Q}}[\epsilon],{\cal Q}[\epsilon]\}\vert \geq 0,
\label{supbound}
\eeq
the equality being saturated if and only if
\beq
\delta_{\epsilon}\Psi = 0 \;\;\; , \;\;\; \Psi \in \{\Psi\}
\label{deluno}
\eeq
\beq
\hat{\cal D}_{i}(\omega,{\cal A})\epsilon = 0
\label{deldos}
\eeq
Since the argument holds for any integration surface $\Sigma$,
this last condition becomes in fact
\beq
\hat{\cal D}_{\mu}(\omega,{\cal A})\epsilon = 0
\label{ndeltacero}
\eeq
We  recognize in (\ref{ndeltacero}) the condition
that $\epsilon$ must satisfy in order to be a Killing spinor.
Moreover, one can see that eqs.(\ref{deluno})-(\ref{ndeltacero})
are the Bogomol'nyi equations of our theory and that
eq.(\ref{supbound}) allows the obtention of a Bogomol'nyi bound.
To see this we shall  establish at this point a link
between   our approach and that of Deser and Teitelboim
\cite{DT} based on
Dirac formalism for
constrained systems. Let us start by recalling that,
it was proven in refs.\cite{DT}-\cite{T} that supergravity charges
obey a supersymmetry algebra at space-like infinity which is
nothing but the usual (global) flat-space algebra. In our notation,
\beq
\{\bar{{\cal Q}}[\epsilon],{\cal Q}[\epsilon]\}\vert =
\opar (\infty)\gamma^{\mu}\epsilon (\infty)P_{\mu} + i
\opar (\infty)Z\epsilon (\infty)
\label{global}
\eeq
where $Z$ is the central charge. Now, it is well-known
that the
central charge in $N=2$ supersymmetric theories, is related
to the topological charge of the
field configuration $T$ \cite{WO}-\cite{ENS}
\beq
Z = iT.
\label{id}
\eeq
Then, for static configurations, eq.(\ref{global}) can be written as
\beq
\{\bar{{\cal Q}}[\epsilon],{\cal Q}[\epsilon]\}\vert =
(M - T)\opar_{\infty}\epsilon_{\infty}\Theta^{2}(R).
\label{global2}
\eeq
which is nothing but our eq.(\ref{lns}) originally obtained
by computing the Poisson bracket through
the formula (\ref{qeqe}).
Using these results and  eq.(\ref{supbound}) one easily writes
a Bogomol'nyi bound for the mass,
\beq
M \geq T.
\label{bou}
\eeq

In view of eqs.(\ref{lns}), (\ref{global})-(\ref{id})
we are lead to identify the topological
charge $T$ of the configuration with the circulation of the vector
${\cal A}_{\mu}$:
\beq
T \equiv \oint_{\partial\Sigma}{\cal A}_{\mu}dx^{\mu}.
\label{tete}
\eeq
Note that eq.(\ref{tete}) does not imply ${\cal A}_{\mu} = A_{\mu}$.
Indeed, they can differ in terms with vanishing
circulation\footnote{For example, in the Abelian Higgs model
coupled to $N=2$ supergravity ,
${\cal A}_{\mu}$ has a contribution proportional to
the Higgs current whose circulation at infinity vanishes
for finite energy configurations \cite{BBS,ENS2}.}.

Taking into account (\ref{bou}), we can identify eqs.(\ref{deluno}) as
the Bogomol'nyi equations of the matter fields. Indeed, given a
particular model, conditions (\ref{deluno}) lead to a first
order differential equation whose solutions satisfy the
second order Euler-Lagrange equations.
Moreover, eq.(\ref{ndeltacero}) is to be
interpreted as the Bogomol'nyi equation of the gravitational
field in the sense that its consistency equation
coincides with Einstein equation
for the gravitational field coupled to matter.

As pointed out in the introduction,
one cannot naively assume the existence
of non-trivial solutions to the Killing spinor equation in view
of the problems posed by conical geometry in three dimensional
spacetime.
To examine in  detail this issue let us
consider a spinor $\eta$ parallel
transported around a closed curve surrounding
the matter sources which, after a gauge transformation, reads:
\beq
\eta(x)\vert_{2\pi} = {\cal P}\exp\left(- \ihalf\oint_{\Gamma}
\omega_{\mu}^0\gamma_0dx^{\mu} + i\frac{\kappa^2}{4}
\oint_{\Gamma}{\cal A}_{\mu}dx^{\mu}\right)\eta(x)\vert_0.
\label{paral2}
\eeq
Then, as a consequence of the fact that the
gravitino has acquired a charge,
a Bohm-Aharonov phase appears, in addition to the
usual phase arising from the
spinorial connection.

Now,
if one assumes, for static configurations,
$\gamma^{0}\eta = \eta$, and then uses eq. (\ref{masa})
together with eq.(\ref{tete}), it is easy to see that
whenever the Bogomol'nyi bound is saturated, the holonomies
in (\ref{paral2}) cancel each other.
That is, unbroken supersymmetries can
be defined over Bogomol'nyi saturated states. It is
worthwhile to remind that, as noted above,
the existence of Killing spinors
implies the existence of non-trivial solutions to Einstein
equations.

Concerning eqs.(\ref{deluno}), we recognize the usual
Bogomol'nyi equations for bosonic matter fields. Let us stress
at this point that solutions of eqs.(\ref{deluno}) break half of
the supersymmetries; indeed, writing the spinor parameter as
\[ \epsilon \equiv \left( \begin{array}{c} \epsilon_+ \\
\epsilon_- \end{array} \right) \]
one can easily see that
\beq
\delta_{\epsilon_+}\Psi = 0 \Longrightarrow
\delta_{\epsilon_-}\Psi \neq 0
\label{sino}
\eeq
for non-trivial configurations. Note that, as pointed out in
\cite{W2} for the Abelian Higgs model \cite{BBS,ENS2},
the possibility of defining global conserved
supercharges for solitonic states that saturate a Bogomol'nyi
bound, does not imply Bose-Fermi degeneracy.

Before illustrating our ideas with the analysis of a
simple model,
we shall descibe as promissed,  how the
Deser-Teitelboim (DT) Hamiltonian approach \cite{DT}-\cite{T}
eliminates
ambiguities afflicting the supercharge algebra, so that
all  the expressions we have used
involving supercharge generators
are well-defined.
Within the DT formalism, one starts from the non-improved
(naive) Hamiltonian $H_0$ and then one adds Lagrange multipliers
for the total energy-momentum,
angular momentum and supercharge.
Then, one  fixes the gauge by means of coordinate conditions that imply a
preferred local time so that space-like surfaces, over whose contour
the generators are defined, become meaningful. One should note
that gauge conditions make the asymptotic form of the Lagrange
multipliers determine the multipliers everywhere.

Since, as noted above, the expression (\ref{qsurf}) for the
supercharge coincides with that arising in pure supergravity,
the ambiguity problem aflicting our supercharge algebra
is just the same as that treated by Deser and Teitelboim \cite{DT}.
That is, if one writes the charge,
within the Hamiltonian approach in terms of the generator of gauge
transformations,

\beq
Q_{H_{0}} = \int d^2x \overline\epsilon \epsilon_{ij}\hat{\nabla}_j
\psi_i
\label{quham}
\eeq
Poisson brackets cannot be calculated
for $\epsilon$
 not vanishing at infinity:
there are surface
terms preventing the naive
evaluation of Poisson brackets.
Within the  DT approach, this problem is solved by
adding to $H_0$  a
boundary terms of the form \cite{T,DT}
\beq
H_s = \overline{\psi}_0(\infty)(\frac{2}{\kappa}
\oint_{\partial\Sigma}\psi_{\mu}dx^{\mu}) = - {\cal
Q}[\psi_{0}(\infty)]
\eeq
Analogous terms have to be added in connection with energy-momentum
and angular momentum but
we will include them since they play no role
in our arguments. Then, the improved Hamiltonian takes the form:
\beq
H_{imp} = {H_0} + H_s
\eeq
Concerning supercharges,
this ammounts to define the supercharge as
\beq
{\cal Q}_{H} = Q_{H_{0}} - \frac{2}{\kappa} \oint \overline\epsilon
\psi_i dx^i
\label{quham2}
\eeq
In this way ${\cal Q}_{H}$
becomes second class
and the supersymmetry algebra can be obtained by
computing Dirac brackets.
Since the boundary term in $H_s$ is the same as that
considered in \cite{DT} for
pure supergravity and $\cal{Q}$ coincides with the
one defined there,
one naturally arrives to Deser-Teitelboim results
forthe models we are discussing.
It becomes
clear within this approach
that the inequality in eq.(\ref{supbound}) is referred to
the ADM mass.

{}~

\noindent{\bf The {CP}$^n$ example}
\vspace{0.2 cm}

We now illustrate  the ideas described
in the precedent section by analysing
the gravitating ${CP}^n$ model.
The $N=2$ supersymmetric
action in three dimensional spacetime can be obtained by
dimen\-sional reduction of the $N=1$ supersymmetric ${CP}^n$ model
in $d=4$ spacetime \cite{CS}.
We will quote here just the bosonic
part of the $N=2$ action
\beq
S = \int d^3x \left( - \frac{1}{4\kappa^2}eR +
\frac{e}{2}(D_{\mu}{\bf z})^*(D^{\mu}{\bf z}) \right)
\label{acsig}
\eeq
where ${\bf z} = (z^1 \ldots z^n)$ are $n$ complex scalar fields,
$e$ is the determinant of the dreibein, $R$ is the scalar
curvature and $D_{\mu} = \dmu - iA_{\mu}$, with $A_{\mu}$
an auxilliary field. The following condition on the scalar
field holds
\beq
|{\bf z}|^2 = \sum_{a}|z^a|^2 = 1
\label{sigcon}
\eeq
together with its corresponding supersymmetric partner.
The set of supersymmetry transformations for
the fermionic partners of those fields appearing in (\ref{acsig})
are
\beq
\delta\psi_{\mu}\vert \equiv \frac{2}{\kappa}\hat{\cal
D}_{\mu}(\omega,A)\vert\epsilon = \frac{2}{\kappa}\left({\cal
D}_{\mu}(\omega)\vert + \frac{\kappa^2}{4}
{\bf z}\dboth_{\mu}{\bf z}^*
\right)\epsilon
\label{sup1}
\eeq
and
\beq
\delta{\bf\chi}\vert = \half\dslash{\bf z}\epsilon
\label{sup2}
\eeq
where ${\bf\chi}$ is the supersymmetric partner of ${\bf z}$
and $\epsilon$ is a complex (Dirac) spinor, reflecting
the existence of an extended
supersymmetry. Eq.(\ref{sup1}) reflects nothing but
the extension of the
supercovariant derivative whenever a vector multiplet is
coupled to the Einstein multiplet as discussed after
eq.(\ref{newcov}) with
\beq
{\cal A}_{\mu} \equiv -i{\bf z}\dboth_{\mu}{\bf z}^*
\label{ident}
\eeq

It is immediate to compute the supercharge
${\cal Q}[\epsilon]$ from the conserved
supercurrent
\beq
{\cal J}^{\mu} = - \half\epsilon^{\mu\nu\rho}\opsi_{\rho}\hat{\cal
D}_{\nu}(\omega,A)\epsilon - \frac{1}{2\sqrt{2}}\bar\chi^a\gamma^{\mu}
\dslash\phi^a\epsilon
\label{curr}
\eeq
which, after Euler-Lagrange equations, can be written
as the surface integral given in (\ref{qsurf}). One can then compute
the supercharge algebra from the circulation of a generalized Nester
form and follow the steps previously described to get, after static
conditions are imposed, the Bogomol'nyi
bound $M \geq |T|$ with $T$ given by:
\beq
T = -i\oint_{\partial\Gamma}{\bf z}\dboth_{\mu}{\bf z}^*dx^{\mu}.
\label{topo}
\eeq
Note that in the present model, ${\cal A}_{\mu}$ coincides with
the vector potential of the topological current.

Following the general prescriptions described above,
it is straigthforward to find the self-duality equation for
the scalar fields, after choosing a spinorial parameter $\epsilon_+$:
\beq
\delta_{\epsilon_+}{\bf\chi}\vert = 0 \Longrightarrow
D_i{\bf z} = i\epsilon_{ij}D_j{\bf z}
\label{sd}
\eeq
as well as the Bogomol'nyi equation for the gravitational field
\beq
\delta_{\epsilon_+}\psi_{\mu}\vert = 0 \Longrightarrow
\left({\cal D}_{\mu}(\omega)\vert + \frac{\kappa^2}{4}
{\bf z}\dboth_{\mu}{\bf z}^*\right)\epsilon_+ = 0
\label{boggrav}
\eeq
whose integrability condition
\beq
\left[{\cal D}_{\mu}(\omega)\vert + \frac{\kappa^2}{4}
{\bf z}\dboth_{\mu}{\bf z}^*,{\cal D}_{\nu}(\omega)\vert +
\frac{\kappa^2}{4}{\bf z}\dboth_{\nu}{\bf z}^*\right]\epsilon_+ = 0
\label{intcond}
\eeq
reproduces the Einstein equation once eq.(\ref{sd}) is imposed.
Had we considered a spinorial parameter $\epsilon_-$, we would
have end with self-gravitating antisolitonic configurations.
Let us end our discussion on the ${CP}^n$ model stressing
that eq.(\ref{boggrav}) admits non-trivial solutions.
It is clear, since the holonomy produced by the conical
nature of spacetime at infinity
is explicitely cancelled by the holonomy of
${\cal A}_{\mu}$ (see eq.(\ref{ident}))
over Bogomol'nyi saturated configurations.

{}~

We conclude this work summarizing its main results. We have proven
on general grounds that Bogomol'nyi bounds and equations in
three dimensional gravitational theories, can be
viewed as the fingerprint of an underlying $N=2$
supersymmetric structure. We have shown that  saturation
of the bound is accompanied by the existence of non-trivial
supercovariantly constant spinors even for
asymptotically conical
spacetime
geometries, due to a cancellation
of holonomies.
Unbroken supercharges can then be defined
for Bogomol'nyi saturated states, this fact leading to the
vanishing of the cosmological constant without having
Bose-Fermi degeneracy \cite{W3,W2}.
The presence of Killing spinors implies the existence of
non-trivial solutions to Einstein equations.
Finally, we have presented
as an example the ${CP}^n$ model which clearly illustrates our
arguments.

{}~

\underline{Acknowledgements}:
This work was
supported in part by  CICBA and CONICET, Argentina and the
Minist\`ere de l'Enseignement Sup\'erieure et de la
Recherche
(France).
F.A.S.~thanks the Laboratoire de Physique Th\'eorique ENSLAPP
for its kind hospitality.


\end{document}